\documentclass[aip,amsmath,amssymb,reprint]{revtex4-1}
\usepackage{graphicx}
\usepackage{dcolumn}
\usepackage{bm}
\usepackage[utf8]{inputenc}
\usepackage[T1]{fontenc}
\usepackage{mathptmx}
\usepackage{color}

\begin{document}
\preprint{AIP/123-QED}

\title{Non-invasive time-sorting in radio-frequency compressed ultrafast electron diffraction}

\author{Lingrong Zhao} \author{Jun Wu} \author{Zhe Wang} \author{Heng Tang} \author{Xiao Zou} \author{Tao Jiang} \author{Pengfei Zhu}
	\affiliation{Key Laboratory for Laser Plasmas (Ministry of Education), School of Physics and Astronomy, Shanghai Jiao Tong University, Shanghai 200240, China}
	\affiliation{Collaborative Innovation Center of IFSA (CICIFSA), Shanghai Jiao Tong University, Shanghai 200240, China}
\author{Dao Xiang} 
	\thanks{Authors to whom correspondence should be addressed: dxiang@sjtu.edu.cn and jzhang1@sjtu.edu.cn}
	\affiliation{Key Laboratory for Laser Plasmas (Ministry of Education), School of Physics and Astronomy, Shanghai Jiao Tong University, Shanghai 200240, China}
	\affiliation{Collaborative Innovation Center of IFSA (CICIFSA), Shanghai Jiao Tong University, Shanghai 200240, China}
	\affiliation{Tsung-Dao Lee Institute, Shanghai 200240, China}
	\affiliation{Zhangjiang Institute for Advanced Study, Shanghai Jiao Tong University, Shanghai 200240, China}

\author{Jie Zhang}
	\thanks{Authors to whom correspondence should be addressed: dxiang@sjtu.edu.cn and jzhang1@sjtu.edu.cn}
	\affiliation{Key Laboratory for Laser Plasmas (Ministry of Education), School of Physics and Astronomy, Shanghai Jiao Tong University, Shanghai 200240, China}
	\affiliation{Collaborative Innovation Center of IFSA (CICIFSA), Shanghai Jiao Tong University, Shanghai 200240, China}

\date{\today}

\begin{abstract}
We demonstrate a non-invasive time-sorting method for ultrafast electron diffraction (UED) experiments with radio-frequency (rf) compressed electron beams. We show that electron beam energy and arrival time at the sample after rf compression are strongly correlated such that the arrival time jitter may be corrected through measurement of the beam energy. The method requires minimal change to the infrastructure of most of the UED machines and is applicable to both keV and MeV UED. In our experiment with $\sim$3 MeV beam, the timing jitter after rf compression is corrected with 35 fs root-mean-square (rms) accuracy, limited by the $3\times10^{-4}$ energy stability. For keV UED with high energy stability, sub-10 fs accuracy in time-sorting should be readily achievable. This time-sorting technique allows us to retrieve the 2.5 THz oscillation related to coherent A$_{1g}$ phonon in laser excited Bismuth film and extends the temporal resolution of UED to a regime far beyond the 100-200 fs rms jitter limitation. 
\end{abstract}

\maketitle

\section{\label{sec:intro}INTRODUCTION}
Ultrafast electron diffraction (UED) has emerged as a powerful tool with high temporal-spatial resolving power, providing direct insight into the structural dynamics of matter \citep{UEDZewail,UEDMiller}. In UED experiments, the dynamics are initiated by an ultrashort pump laser and probed by a delayed electron pulse. By recording diffraction patterns at a series of pump-probe delays, it is possible to retrieve atomic changes following laser excitation. This approach has been applied to study dynamics in phase transition \cite{AlMelt2003,baum20074d, gedik2007nonequilibrium, Miller2009SciWDM, Miller2009NatAcc, Miller2010NatCDW, Haupt2016PRLCDW, Mo2018AuMelting}, to reveal transient states \cite{raman2008direct, Siwick2014SciVO2, Kogar2020NatPhyLaTe3, zhou2021nonequilibrium,duan2021optical}, and to visualize molecular dynamics \cite{ihee2001direct,JYang2016PRL, yang2018imaging, wolf2019photochemical}. With the advent of femtosecond laser, the time resolution of UED is primarily limited by the pulse width and timing jitter of the electron bunch. Coulomb repulsion is the main effect that accounts for lengthening of electron pulse width as electron beam propagates from the source to the sample. To circumvent this limitation, many efforts have been made in the past two decades, e.g. reducing the propagation distance \cite{AlMelt2003,baum20074d}, reducing the bunch charge to single electron \cite{fill2006sub,veisz2007hybrid,PNAS2010single}, increasing the electron energy to relativistic regime \cite{SLAC2006MeV, Tsinghua2009MeV, UCLA2010MeV, Osaka2011MeV, SJTU2014MeV, PF2015MeV, SLAC2015MeV, Desy2015MeV}, and compressing the beam with rf buncher\cite{keV2007SH100fs, van2010compression, gao2012full,2012APLRFued, chatelain2014coherent, Maxson2017sub10fs, zandi2017high, Zhao2018PRX}, THz buncher\cite{kealhofer2016all, Zhao2020THz, SLAC2020THz} and double bend achromat \cite{PRL2020DBA, NP2020KoreanDBA}.

The most widely used bunch compression method for UED is velocity bunching with an rf buncher \cite{keV2007SH100fs, van2010compression, gao2012full,2012APLRFued, chatelain2014coherent, Maxson2017sub10fs, zandi2017high, Zhao2018PRX}, and the shortest bunch duration is below 10 fs rms with bunch charge on the order of 10 fC for MeV beams \cite{Maxson2017sub10fs, Zhao2018PRX}. However, previous experiments have shown that with rf buncher, the electron beam pulse width is reduced at the cost of increasing the timing jitter which is typically measured to be about 100-200 fs rms \cite{2012APLRFued,gao2012full,zandi2017high, Zhao2018PRX}. The primary cause for this timing jitter is the phase jitter between the laser oscillator and rf electronics. Specifically, the phase jitter in the rf cavity leads to beam energy jitter which is further converted into timing jitter at the sample after passing through a drift with longitudinal dispersion. While the timing jitter can be measured with a THz deflector \cite{Zhao2018PRX,Li2019PRAB, zhao2019THzOsci}, and very recently a THz-streaking based method \cite{othman2021visualizing} has been used to correct the jitter, the method is invasive and it nonetheless introduces changes to the diffraction pattern which may be difficult to differentiate from those by laser excitation.  

In this paper, we demonstrate a non-invasive time-sorting method to record the arrival time of rf compressed relativistic electron beam through measurement of the beam energy of the un-diffracted beam. Our measurements show that after rf compression the beam arrival time at the sample is strongly correlated with beam energy. Because the detector has a central hole to allow the un-diffracted beam to path through, it is straightforward to simultaneously measure both the diffraction pattern and beam arrival time with a downstream energy spectrometer. This method has been used to retrieve the A$_{1g}$ phonon oscillation in laser excited Bismuth. In contrast, no oscillation has been observed in the raw data without jitter correction. This measure-and-sort method is applicable to both keV and MeV UED and can be used to correct both short-term timing jitter and long-term timing drift in rf-compressed UED. In our experiment, the timing jitter of the rf compressed beam is corrected with 35 fs rms accuracy, limited by the energy stability of the electron beam before the buncher. Since keV UED has high stability in beam energy, an accuracy of a few femtosecond should be readily achievable with this time-sorting method. 

\section{Principle of bunch compression and source of timing jitter}
In this section, we briefly discuss the physics behind the time-sorting technique. For simplicity, we consider a MeV electron beam with negligible energy chirp (correlation between an electron's energy and longitudinal position). In the most common velocity bunching scheme, the beam is sent through an rf buncher cavity at the zero-crossing phase. Because the bunch head is decelerated and the bunch tail is accelerated, the beam ends up with a negative energy chirp (bunch head has lower energy than bunch tail) at the exit of the buncher. The energy chirp can be calculated as,
\begin{equation}
h=\dfrac{d\delta}{dz}=-\dfrac{2\pi V}{E \lambda}
\end{equation}
where $E$ is the beam energy, $V$ and $\lambda$ are the voltage and wavelength of the rf buncher, respectively. The longitudinal dispersion of a drift with length $L$ is $R_{56}=cdt/d\delta\approx -L/\gamma^2$, where $\gamma$ is the Lorentz factor of the electron. By matching the longitudinal dispersion with the beam energy chirp such that $hR_{56}=1$, the beam will be fully compressed after a drift. In this case, the electrons with higher energy at the bunch tail exactly catch up with the lower energy electrons at the bunch head after the drift. 

Ideally, the electron beam passes through the buncher at zero-crossing phase such that the energy chirp is imprinted without changing the beam centroid energy. In realistic cases, the phase jitter of the rf field leads to change of the beam centroid energy. Taken the phase seen by the electron beam to be $\Delta \phi$ (much smaller than 1), at the buncher exit the change of beam centroid energy is $\Delta\delta \approx (V/E)\Delta \phi$. After the drift, this energy change will result in a time-of-flight change of $\Delta t=\Delta\delta R_{56}/c$. Under the full compression condition, the time-of-flight change is simplified to $\Delta t=-\Delta\phi_t$, where $\Delta\phi_t=\Delta\phi/(kc)$ is the normalized phase jitter that has the same unit as time and $k$ is the wave number of the rf and $c$ is speed of light. From discussions above, one can see that the timing jitter is correlated with the energy jitter with the coefficient being the longitudinal dispersion of the drift. Therefore, the timing jitter may be determined by measuring the beam energy after rf compression.  

Simulation with General Particle Tracer code \cite{GPT} is used to verify this dependence and to also take into account the effect of rf amplitude and phase jitter in both the photocathode rf gun and buncher cavity. Following our experimental setup, in this simulation the electron beam is produced in a 1.6 cell s-band (2856 MHz) photocathode rf gun and compressed in a 5 cell c-band (5712 MHz) buncher. The accelerating gradient and launching phase in the gun are set to 58 MV/m and 38 degrees, respectively. The phase is chosen to minimize the timing jitter at the entrance to the buncher \cite{LRK2009NIM}. The buncher is located at 0.7 m downstream of the gun and the bunching voltage is set to 0.9 MV which resulted in full compression after a drift of 1.0 m. The amplitude jitter and phase jitter in the two rf cavities are both set to 0.03\% and 150 fs, respectively. In the simulation, the kinetic energy of the electron beam is 2.56 MeV and the corresponding $R_{56}$ from the exit of the rf buncher to the sample is -2.77 cm (or -92.3 ps). 

10000 runs of simulation with single electron are performed and the electron's arrival time and energy at the sample are shown in Fig.~1(a) (blue dots) where one can see that the arrival time of the electrons is linearly correlated with their energy. The coefficient is equal to the longitudinal dispersion of the drift. The rms timing jitter is about 160 fs rms, as shown in Fig.~1(b). After removing the linear term, the residual timing jitter (magenta dots) which can not be corrected by measuring the beam energy is about 30 fs rms. Analysis shows that the residual jitter is comparable to the beam timing jitter at the entrance to the rf buncher which strongly depends on rf amplitude jitter in the gun. The simulation was repeated by reducing the rf amplitude jitter in the gun to 0.01\% while keeping other parameters unchanged, and the simulation results are shown in Fig.~1(c) and Fig.~1(d) where one can see that the residual jitter is similarly reduced by 3 times to about 10 fs rms. In keV UED with a DC electron gun, relative ripple of the accelerating voltage is typically less than $10^{-5}$ and the beam arrival time should be determined with an accuracy well below 10 fs using this method.
\begin{figure}
\includegraphics[width =8.6cm]{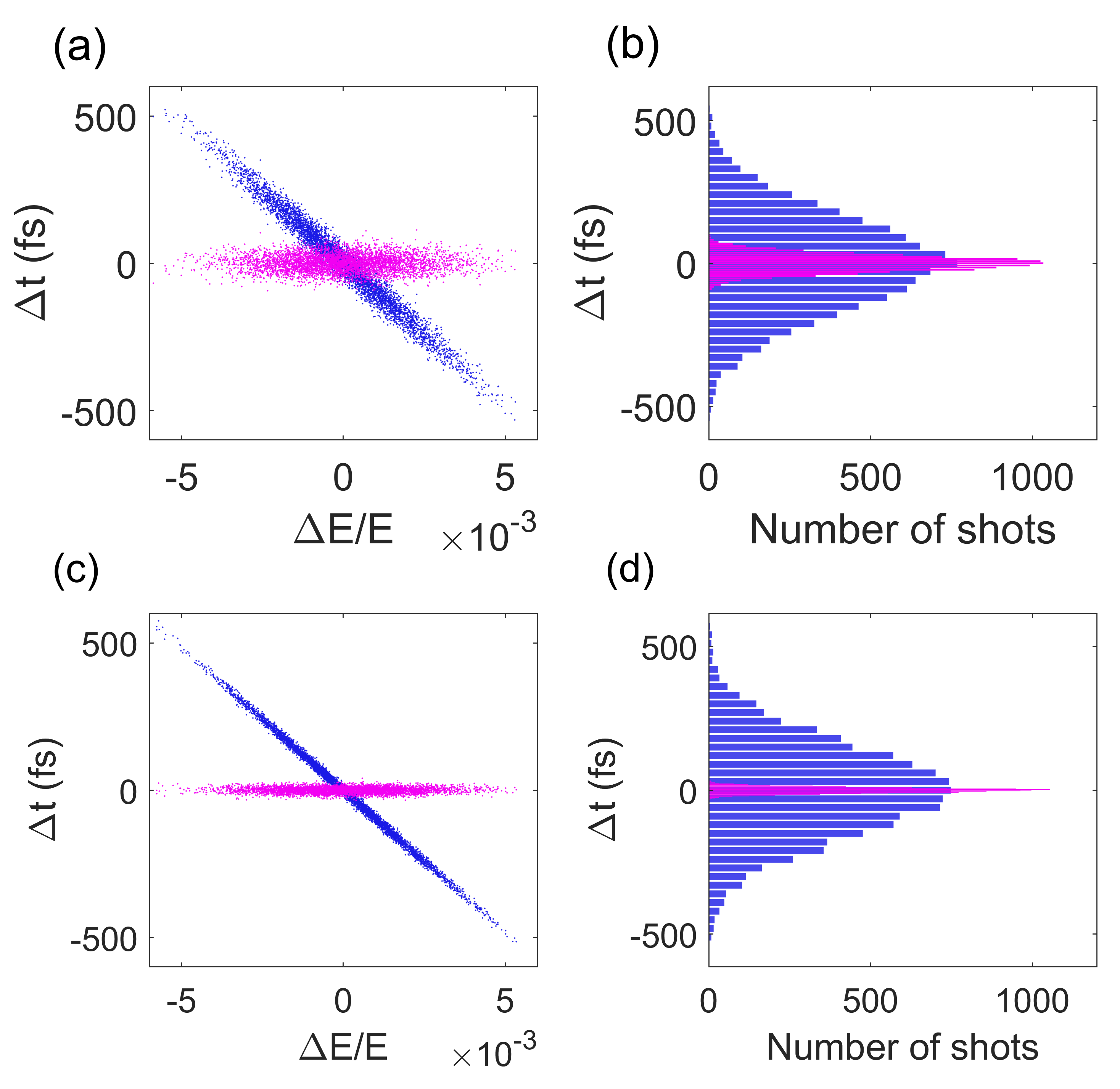}
\caption{\label{fig:1} Simulated electron timing jitter and energy jitter at the sample with the rf amplitude jitter  set to 0.03\% (a) and 0.01\% (c). Panel (b) and (d) are the corresponding distributions of the timing jitter in (a) and (c) respectively.}
\end{figure}
\begin{figure*}
\includegraphics[width =16cm]{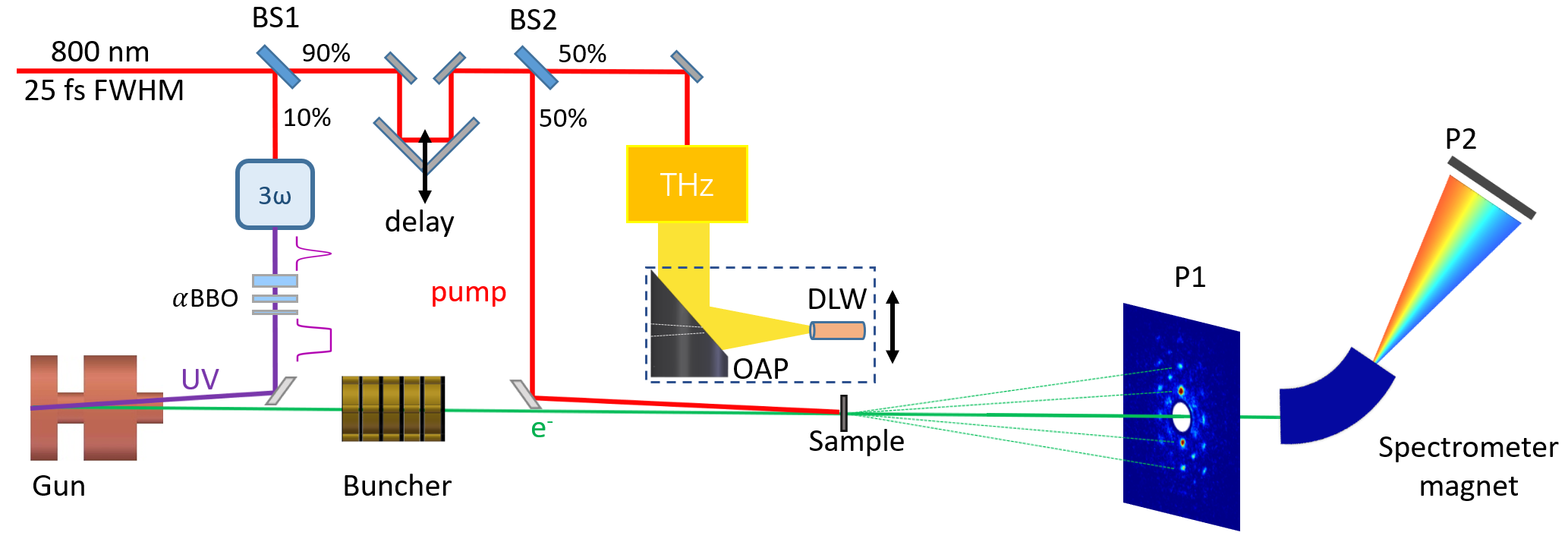}
\caption{\label{fig:2}(a) Schematic of the experimental setup. The 800 nm laser is split into three pulses with one pulse for producing electron beam, the second pulse to excite dynamics in the sample, and the third pulse for producing THz radiation. The electron beam with about 100 fC charge is compressed with an rf buncher. The pulse duration and time jitter is measured with THz streaking in a dielectric-lined waveguide. The diffraction pattern is measured with screen P1 which has a hole to allow the un-diffracted beam to pass through. The distance between the sample and screen P1 is about 1.8 m. A dipole magnet downstream of P1 is used to measure the energy of the un-diffracted beam at screen P2. BS for beam splitter.}
\end{figure*}
\section{Electron beam compression and characterization}
The schematic of our experiment to demonstrate non-invasive time-sorting is shown in Fig.~2. For producing electron bunches with sufficient charge to generate high quality single shot diffraction patterns, a set of $\alpha$BBO crystals are used to shape the temporal profile of the UV pulse from about 50 fs Gaussian distribution to $ \sim $2 ps FWHM flat-top distribution. The $\sim$2.6 MeV electron beam is compressed by a C-band (5712 MHz) rf buncher cavity. The pulse duration and arrival time jitter of the electron beam at the sample is characterized by a THz deflector \cite{zhao2019THzOsci}.  

The THz pulse is generated through optical rectification in LiNbO$_3$ crystal \cite{Hebling2002TPF}. An off axis parabolic (OAP) mirror collects and focuses the THz radiation with vertical polarization into a dielectric-lined waveguide (DLW) where HEM11 mode is excited \cite{gallot2000THzwaveguide}. The electron beam in the THz deflector receives a vertical time-dependent angular streaking, which maps its time information into spatial distribution on screen P1.

We first measured the streaking deflectogram with a short electron beam ($\sim$100 fs rms) by removing the BBO crystals. After both the spatial and temporal overlap between the THz and electron beam are optimized, the beam deflectogram [Fig.~3(a)] is measured with the timing of the THz beam varied in 30 fs step. The maximal streaking ramp (around t = 2.4 ps region in Fig.~3(a)) is found to be about 6.0 $\mu$rad/fs. The dynamic range of this measurement where the rate of angular change is approximately linear is about 400 fs. The accuracy of the arrival time measurement is mainly affected by the fluctuation of the centroid divergence of the electron beam, leading to temporal offset in the measurement. In this experiment, the beam centroid fluctuation on P1 with THz off is measured to be about 9.0 $\mu$rad, corresponding to an accuracy of about 1.5 fs for the jitter determination. The transverse beam size at screen P1 was measured to be about 190 $\mu$m with the THz off [Fig.~3(b)], and the temporal resolution in beam temporal profile measurement is estimated to be about 18 fs. 

After the streaking deflectogram is obtained, the BBO crystals are then inserted back to produce a 2-ps flat-top UV pulse and in this case an electron bunch with similar pulse width is produced. With the rf buncher off, the streaked beam has a double-horn distribution [Fig.~3(c)], indicating that the electron pulse extends over at least half of the period of the THz streaking field. We then turned on the C-band buncher for compressing the beam. The buncher voltage is varied until the smallest streaked beam size on screen P1 is obtained. In this case, the beam reaches shortest bunch length at the sample and the streaked beam is shown in Fig.~3(d). 

The bunch length at full compression can be estimated by an analysis of the vertical projections with and without THz streaking. The red circles and blue dash curve in Fig.~3(e) are the profiles of the compressed beam with and without THz streaking, respectively. By fitting the raw streaked distribution with a Gaussian function (black), the bunch length is estimated to be about 30 fs rms. After subtracting the contribution from the intrinsic beam size, the bunch length after deconvolution is estimated to be 25 fs rms.

\begin{figure}
\includegraphics[width = 8.5cm]{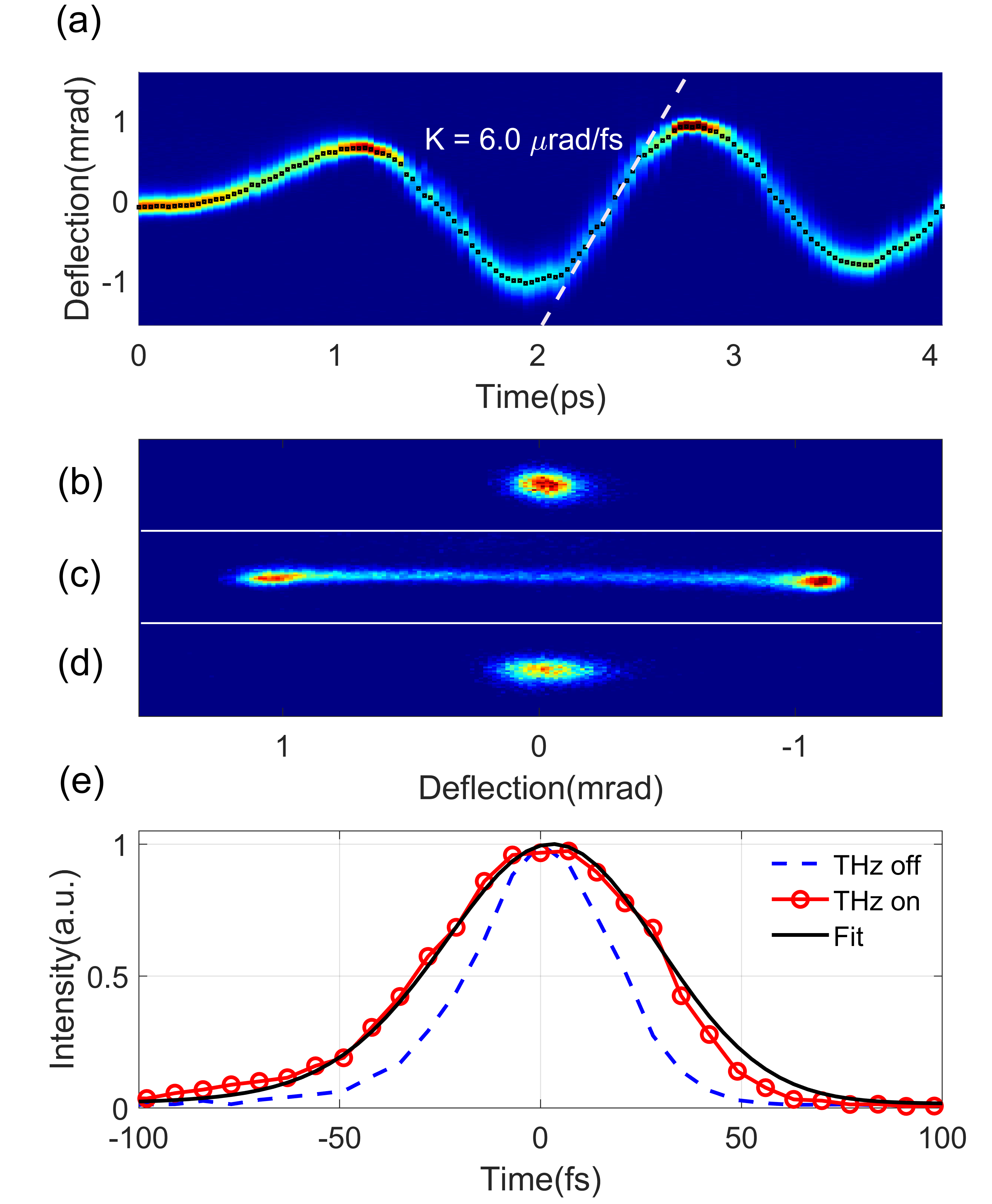}
\caption{\label{fig:3} (a) Streaking deflectogram with maximal streaking rate of 6.0 $\mu$rad/fs. (b) Raw distribution of the electron beam with the THz off. Panel (c) and (d) are distributions of streaked beam before (c) and after (d) rf compression. (e) Fit to the experimental trace to estimate the bunch length at full compression.}
\end{figure}

\section{Timing jitter measurement} 
Because of the 100-200 fs rf phase jitter, the electron beam is compressed at the cost of increasing the timing jitter to a similar level. Under full compression condition, 50 consecutive measurements of THz-streaked beam profiles are shown in Fig.~4(a). For comparison, we also included consecutive measurements of the beam profile with THz off (the first 10 shots). The arrival time of the electron beam at the sample is determined from the centroid of the streaked profile. The timing jitter at full compression collected over 2500 shots is estimated to be about 140 fs (rms), as shown in Fig.~4(b). 

Since the path length from the buncher to the sample is fixed, the change in time-of-flight is essentially related to the change of velocity which strongly depends on beam energy. In order to show the correlation between the beam energy and arrival time and estimate the precision of the proposed time-sorting method, we conducted a separate measurement at screen P2 downstream of the energy spectrometer. Because the electron beam is bent in horizontal direction by the energy spectrometer and streaked in vertical direction by the THz deflector, the horizontal axis on screen P2 becomes the energy axis and the vertical axis becomes the time axis. This allows us to measure both the arrival time and beam energy simultaneously. The measured correlation between the arrival time and centroid energy of the beam is shown in Fig.~4(c) (blue dots), where one can see that the beam timing jitter is indeed linearly correlated with the beam energy jitter, i.e., $\Delta$t = R$\times\Delta$E/E, with $R$ determined to be about -93 ps, in good agreement with the value of the longitudinal dispersion of the drift. After removing the linear term, the residual timing jitter that limits the accuracy of jitter correction is about 35 fs rms as shown in Fig.~4(d). With this jitter correction, the temporal resolution in this rf-compressed UED has been improved from about 150 fs to about 50 fs rms, comparable to that achieved in a state-of-the-art double bend achromat based UED \cite{PRL2020DBA, NP2020KoreanDBA}, yet with greatly simplified infrastructure. The accuracy in jitter correction is mainly limited by the rf amplitude stability of the photocathode rf gun and sub-10 fs correction accuracy should be readily achievable for keV UED with greatly improved beam energy stability. 

\begin{figure}
\includegraphics[width = 8.6cm]{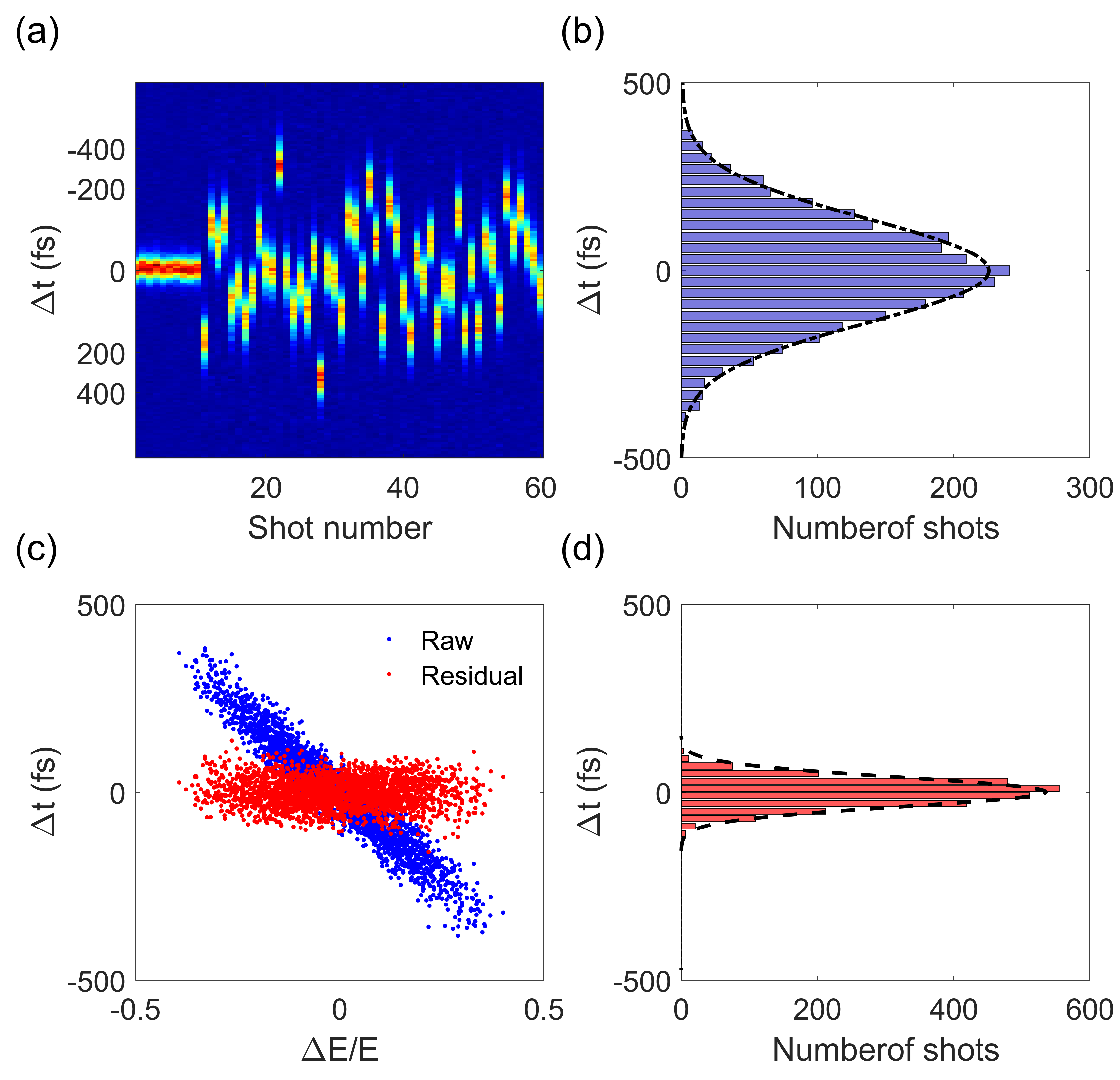}
\caption{\label{fig:4} (a) Consecutive measurement of beam arrival time with THz streaking. (b) Distribution of the electron beam arrival time collected over 2500 shots. A Gaussian fit to the distribution yields a timing jitter of about 140 fs (rms) between the electron beam and THz pulse. (c) Correlation of the beam energy and arrival time. (d) Distribution of the residual jitter after removing the linear term.}
\end{figure}

\section{Non-invasive jitter correction in UED}
\begin{figure*}
\includegraphics[width =17.5cm]{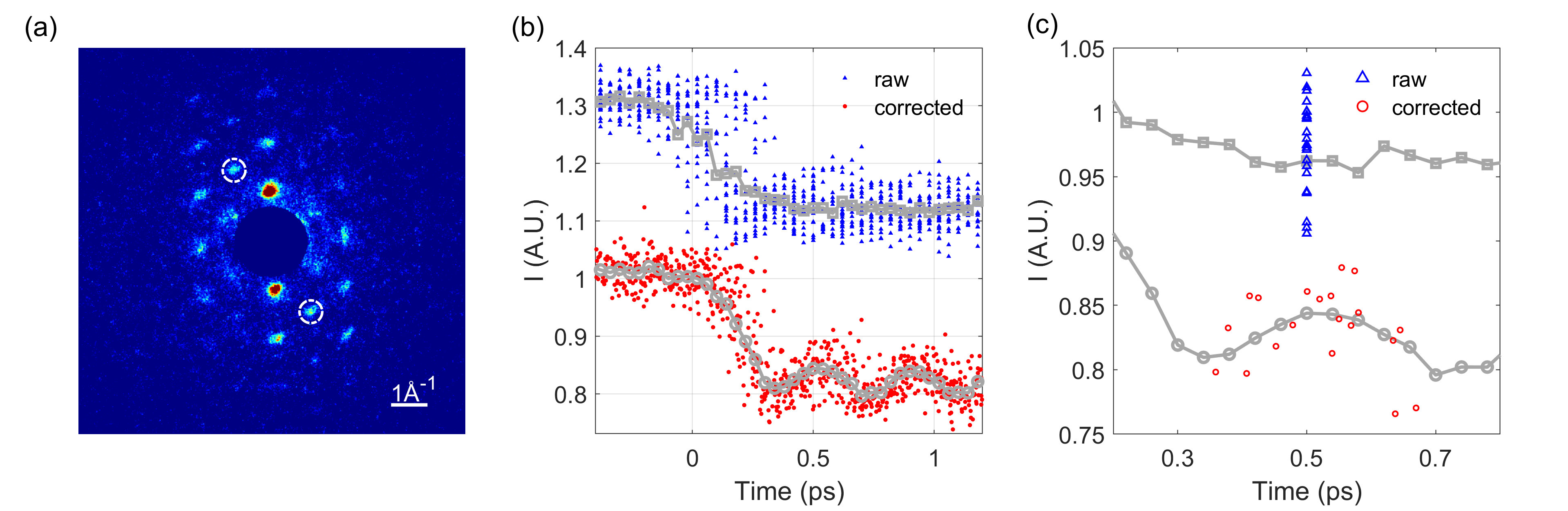}
\caption{\label{fig:5} (a) Single shot diffraction pattern of the single crystal Bismuth.(b) Measured evolution of the Bragg reflection highlighted in (a) by white dashed circle with and without correction of the time jitter. (c) 20 selected data points at 0.5 ps time delay before  timing correction (blue triangle) and the corresponding distribution after timing correction (red circle).}
\end{figure*}

The feasibility of this non-invasive time-sorting technique is demonstrated in a pump-probe experiment to measure the A$_{1g}$ phonon dynamics of Bismuth excited by a 800 nm laser. For pump-probe experiment, after the full compression condition is achieved at the sample, the OAP and DLW are removed from the beam path with an in-vacuum stage and the sample is inserted into the beam path. A pump laser pulse is used to excite the ultrafast structural dynamics in the sample. The diffraction pattern and beam energy of the unscattered beam is synchronously recorded on screen P1 and P2, respectively. It should be noted that screen P1 has a 2-mm-diameter hole in the center to let the un-diffracted beam to pass through. 

A 30-nm-thick (110) oriented Bismuth film is grown by molecular beam epitaxy on KCl(100) substrate \cite{Moriena2012Bi110}. With femtosecond photoexcitation of charge carriers, coherent A$_{1g}$ optical phonon, corresponding to atomic vibrations parallel to the trigonal axis of the rhombohedral unit cell of Bismuth is produced \cite{Cheng1990BiA1g}. The atomic vibration leads to oscillation in diffraction intensity which has been widely used to test the temporal resolution of the instrument \cite{PRL2007HXCOP, NP2013fsHXL}. To measure this structural change the sample is rotated by approximately 37 degrees with respect to normal incidence. Representative single shot diffraction pattern obtained with our rf compressed beam is shown in Fig.~5(a). The Bragg reflections sensitive to A$_{1g}$ phonon are marked by white dashed circles. A 800 nm, 25 fs FWHM laser with a fluence of about 1.5 mJ/cm$^2$ is used to excite the dynamics. The time delay of the pump laser and electron beam is controlled by a delay stage with step size of 40 fs. Diffraction pattern and energy of the unscattered electron beam are collected synchronously at 10 Hz, limited by the frame rate of our CCD camera. In this experiment, 800 diffraction patterns (20 diffraction patterns for each time step) are collected and the measured intensity evolution of the highlighted Bragg reflection before and after correcting the time jitter are shown in Fig.~5(b). The A$_{1g}$ coherent phonon oscillation at about 2.5 THz is clearly resolved after jitter correction while no oscillation is observed for the raw data without jitter correction. Fig.~5(c) shows the process of jitter correction for the 20 data points collected at time delay of 0.5 ps. Before correction (blue triangle), the data points have different intensity, but are assigned with the same delay time. After correcting jitter based on the measured beam energy for each data, the corrected data points (red circle) have different delay time (diffaction intensity for each data is not changed in the process) and now the data has a better match to the A$_{1g}$ coherent phonon oscillation. Similar procedures are used to re-sort the data at other delay time and the final results are shown in Fig.~5(b).

\section{CONCLUSIONS}
In conclusion, we have demonstrated a non-invasive method for time-stamping of rf compressed electron beams. The method has been used in a pump-probe experiment to reveal coherent oscillation of A$_{1g}$ phonon mode of Bismuth that is otherwise smeared out by timing jitter in the raw data. With this method, the time resolution of rf-compressed UED is greatly improved. This method requires minimal change (e.g. a bending magnet) to the infrastructure of most of the UED machines and is applicable to both keV and MeV UED. In our experiment, the timing jitter after rf compression is corrected with 35 fs accuracy, limited by the $3\times10^{-4}$ stability of the rf amplitude in the gun. For keV UED with greatly improved energy stability, sub-10 fs accuracy in time-sorting should be readily achievable. It should be noted that this time-sorting method has large dynamic range and picosecond-level jitter may be corrected with similar accuracy. Therefore, it may be used to correct long-term timing drift in rf-compressed UED as well. We expect this method to have a strong impact in rf-compressed UED where the enhanced resolution by jitter correction may open up new opportunities in ultrafast science.

\begin{acknowledgments}
This work was supported by the National Natural Science Foundation of China (Grants No. 11925505, 11504232, 12005132 and 11721091), China National Postdoctoral Program for Innovative Talents (No. BX20200220), and the office of Science and Technology, Shanghai Municipal Government (No. 16DZ2260200 and 18JC1410700). 

The data that support the findings of this study are available from the corresponding author upon reasonable request.
\end{acknowledgments}

\nocite{*}
\bibliography{ref}
\end{document}